\documentclass[12pt]{article}
\usepackage{geometry}
\usepackage{graphicx}
\usepackage{pgfplots}
\usepackage{tikz}
\usepackage{forest}
\useforestlibrary{linguistics} 
\usepackage{amssymb}
\usepackage{epstopdf}
\usepackage{amsthm}
\usepackage{amsmath}
\usepackage[retainorgcmds]{IEEEtrantools}
\usepackage[utf8]{inputenc}
\usepackage{enumerate}
\usepackage{mathtools}
\usepackage{units}
\usepackage[comma,authoryear]{natbib}
\usepackage{hyperref}
\usepackage{setspace}
\usepackage{textgreek}

\usepackage{natbib}
\usepackage{url}
\makeatletter
\def\url@leostyle{%
    \def\UrlFont{\sf}}{\def\UrlFont{\small\ttfamily}}
\makeatother
\title{Justifying the Norms of Inductive Inference}

\begin{document}
\author{Olav Benjamin Vassend}

\def\citeapos#1{\citeauthor{#1}'s (\citeyear{#1})}

\bibliographystyle{chicago}

\numberwithin{equation}{section}

\maketitle

\newcommand{\E}{\mathrm{E}}
\newcommand{\Ev}{\mathrm{Ev}}
\newcommand{\Max}{\mathrm{Max}}
\newcommand{\PredAcc}{\mathrm{PredAcc}}

\begin{abstract}
Bayesian inference is limited in scope because it cannot be applied in idealized contexts where none of the hypotheses under consideration is true and because it is committed to always using the likelihood as a measure of evidential favoring, even when that is inappropriate. The purpose of this paper is to study inductive inference in a very general setting where finding the truth is not necessarily the goal and where the measure of evidential favoring is not necessarily the likelihood. I use an accuracy argument to argue for probabilism and I develop a new kind of argument to argue for two general updating rules, both of which are reasonable in different contexts. One of the updating rules has standard Bayesian updating, \citeapos{bissiri2016} general Bayesian updating, \citeapos{douven2016} IBE-based updating, and \citeapos{vassend2019b} quasi-Bayesian updating as special cases. The other updating rule is novel.
\end{abstract}

\tableofcontents

\section{Introduction}

Bayesians hold that inductive inference requires two ingredients. First, a prior probability function defined on the hypotheses under consideration. Second, a likelihood function, which assigns a probability to the evidence conditional on each hypothesis.  Intuitively, the prior probability assigned to a hypotheses represents how plausible it is that the hypothesis is true before the evidence has been taken into account. The likelihood, on the other hand, is a measure of evidential favoring: if $H_1$'s likelihood on the evidence is greater than $H_2$'s likelihood on the same evidence, then the evidence favors $H_1$ over $H_2$. Given a prior and likelihood, Bayesians hold that the prior probability of each hypothesis should be updated to a posterior probability through the use of Bayes's formula, so that the posterior probability of $H$ is proportional to the prior probability of $H$ multiplied by its likelihood.

Bayesianism has become the most common formal framework used by philosophers of science to study scientific methodology, and it is also an influential framework for statistical inference. But it rests on an assumption that is often violated in scientific practice, namely that one of the hypotheses under consideration is true.\footnote{This limitation is well known, but often ignored. For discussion of the problem, see, e.g. \cite{box1980}; \cite{bernardosmith1994}; \cite{forstersober1994}; \cite{forster1995}; \cite{key1999}; \cite{shaffer2001}; \cite{sprenger2009}; \cite{gelmanshalizi2013}; \cite{vassend2019a}; \cite{walker2013}; and \cite{sprenger2017}.} Suppose none of the hypotheses under consideration is true, so that the goal is instead to find the hypothesis that is -- in some sense -- best. Depending on what is meant by ``best,'' the likelihood may not be an appropriate measure of evidential favoring. For example, suppose the goal is to identify the hypothesis whose expected maximal prediction error on future data is as low as possible. Then, as \cite{vassend2019b} shows, the likelihood is not an appropriate measure of evidential favoring because the hypothesis that has the best likelihood score on the evidence will in general not be the hypothesis that has the lowest expected maximal prediction error on future data. In this context, a more reasonable measure of evidential favoring may be one according to which the evidence favors $H_1$ over $H_2$ if and only if $H_1$'s maximal prediction error on the evidence is lower than $H_2$'s maximal prediction error on the evidence. The fact that Bayesianism is tied to using the likelihood as a measure of evidential favoring is therefore a limitation of the framework. 

The goal of this paper is to study inductive inference in a very general setting. Suppose our goal is to identify the best hypothesis $H$ (where ``best'' does not necessarily mean ``true''). Let $p$ be a function that assigns a number between 0 and 1 (inclusive) to each hypothesis, such that $p(H)$ is interpreted as representing a prior judgment of how plausible it is that $H$ is best (in the relevant sense) out of the hypotheses under consideration. In the rest of the paper, I will refer to any such function as a ``credibility function''. Suppose, moreover, that $\Ev[E|H]$ is an evidential measure that is sensible given the purpose at hand. Then the questions to consider are as follows: (1) What norms should $p$ obey? (2) How should $p(H)$ and $\Ev[E|H]$ be combined in order to produce a posterior score $p_E(H)$ that represents how plausible it is that $H$ is best in light of $E$ and the prior information? 

As we will see, one of the standard Bayesian arguments for probabilism generalizes, so that -- given widely applicable conditions -- $p$ and $p_E$ ought to be probability functions. The more interesting results concern updating. I will show that, depending on what the goal is, the prior probability function and evidential measure should be combined in one of the following two ways in order to produce a posterior probability:

\begin{quote} \textbf{Inferential updating.} Given evidential measure $\Ev$ and prior probability function $p$, update $p$ to the posterior $p_E$ by way of the following formula:

\[ 
p_E(H) = \frac{\Ev{[E | H]}p(H)}{\sum_i \Ev{[E | H_i]}p(H_i)} 
\]

\end{quote}

\begin{quote} \textbf{Predictive updating.} Given evidential measure $\Ev$ and prior probability function $p$, update $p$ to the posterior $p_E$ by way of the following procedure:

\begin{quote} \textbf{Step 1.} For each $i$, calculate $q(H_i) =  p(H_i) + \Ev[E|H_i]$. 

\textbf{Step 2.} Transform $q$ to $p_E$ as follows: for each $i$, $p_E(H_i) = 0$ or $p_E(H_i) =  q(H_i) + d$, where $d$ is the unique number such that $d$ is minimal and, for all $i$, $p_E(H_i) \ge 0$ and $\sum_i {p_E(H_i) = 1}$.  \end{quote} 

\end{quote}

The justification for the names of the two updating procedures will become clearer later. Inferential updating is clearly a generalization of Bayesian updating. Indeed Bayesian updating is just inferential updating with the likelihood used as the measure of evidential favoring.\footnote{Predictive updating, on the other hand, may remind the reader of the alternative to Jeffrey conditionalization derived by \cite{leitgeb2010}. The two rules do indeed share several features in common, although they are also importantly different. In fact, it is possible to derive a special case of predictive updating by using a proof strategy that resembles the one in  \cite{leitgeb2010}.} What separates inferential updating from predictive updating is the former rule's commitment to \emph{Regularity}: inferential updating will never assign a probability of 0 to any hypothesis, whereas predictive updating typically will. In Section 4, we'll see that a commitment to Regularity is sometimes reasonable and sometimes not.

The plan for the rest of the paper is as follows. In Section 2, I sketch an argument for why any credibility function ought to be probabilistic, regardless of whether the goal is truth or something else. Since the argument is a straightforward adaptation of \citeapos{pettigrew2016} accuracy argument for probabilism, the section is brief. In Section 3, I give characterizations of inferential and predictive updating from a set of plausible assumptions. The strategy is to divide inductive updating into two steps: in the first step, the prior plausibility of a hypothesis is combined with the hypothesis's score on the evidence according to some measure of evidential favoring in order to produce a posterior score. In the second step, the posterior scores are normalized so that they are probabilistic. As we'll see, the requirement that the combination step and normalization step commute in certain desirable ways, together with a few other plausible assumptions, result in the conclusion that the combination step and normalization step must both be either multiplicative or additive. The characterizations of inferential and predictive updating are then just a few short steps away. I end the paper with a discussion of inferential and predictive updating, including their relationship to each other and to other updating rules.

\section{Why credibility functions should be probabilistic}

Before we can show that credibility functions ought to be probabilistic, we need to get clearer on what this claim amounts to. Let \textbf{H} be a set of hypotheses and suppose the goal is to identify the hypothesis in \textbf{H} that is best rather than true (where ``best'' can mean anything we like). One complication that arises when ``true'' is replaced by ``best'' is that whereas there is only one true hypotheses, there may be several that are best.\footnote{I thank X for pointing this out to me.} For example, if ``best'' means ``having a minimal maximum expected prediction error,'' then there may be several hypotheses that are tied for best. Note, however, that this is more a theoretical possibility than a practical one, since it is quite unlikely that multiple hypotheses would have (say) exactly the same predictive accuracy score, especially if the number of hypotheses is large. I will henceforth assume that at most one hypothesis out of the hypotheses under consideration is best. Note that if we make this assumption, then the hypotheses will also be mutually exclusive in the sense that in any subset of hypotheses at most one hypothesis can be best. 

Another theoretical possibility is that none of the hypotheses under consideration is best. This can, for example, happen if the hypothesis space is infinite and does not contain a single best hypothesis, but rather an infinite sequence of hypotheses in ascending order of goodness.\footnote{I thank a referee for pointing out this possibility.} To preclude this possibility, we must also assume that at least one of the hypotheses under consideration is best. 

Provided we make the above assumptions (i.e. that exactly one of the hypotheses in \textbf{H} is best), then there is nothing mathematically or philosophically that prevents us from treating \textbf{H} as a sample space. I.e.  \textbf{H} consists of hypotheses that are exhaustive in the sense that one of the hypotheses is best and mutually exclusive in the sense that at most one of the hypotheses is best in any collection of hypotheses. Note also that there is a natural $\sigma$-algebra on \textbf{H}. More precisely, union (or disjunction) and intersection (or conjunction) are defined in the normal way, the identity element for conjunction (i.e. the top element of the algebra) is \textbf{H}, and the complement (negation) of any set $A$ formed through unions and intersections of subsets of \textbf{H} is defined in the following way: $\neg{A} := \textbf{H} - A$. The main difference from the definition given in most philosophical treatments of Bayesianism is that the top element is now \textbf{H} rather than the tautology. This makes a big interpretive difference, but no difference to the mathematics.

Given the above set-up, we can now define what it means for a function on the algebra, $\textbf{H}^{*}$, generated by \textbf{H} to be probabilistic in the following way:

\begin{quote}

\textbf{Probability axioms.} A function $p$ defined on $\textbf{H}^{*}$ is probabilistic if and only if it satisfies the following requirements:

\begin{quote}

1. $p(\textbf{H})=1$.

2. $p(A) \ge 0$ for all subsets $A$ of  $\textbf{H}^{*}$.

3. $p(A \vee B) = p(A) + p(B) - p(A \& B)$, for all subsets $A$ and $B$ of $\textbf{H}^{*}$.

\end{quote}

\end{quote}

Note that credibility functions automatically satisfy 2 since we have defined them to have a range between 0 and 1, so the real question is whether they ought to satisfy 1 and 3. One of the standard arguments for why regular credence functions (or degrees of belief) ought to be probabilistic is the accuracy argument (\cite{joyce1998}, \cite{joyce2009}, \cite{pettigrew2016}, \cite{predd2009}). Briefly, the argument is as follows:\footnote{There are several versions of the argument; here, I present a variant of \citeapos{pettigrew2016} version.} the ideal credence function to have is the function that assigns 1 to the hypothesis that is true and 0 to all hypotheses that are false. Suppose now that we have a divergence measure (satisfying certain reasonable properties) that quantifies the distance between the ideal function and any other candidate credence function. It can then be shown that any credence function that is not probabilistic will be dominated by some probabilistic function in the sense that the probabilistic function will be guaranteed to have a smaller divergence from the ideal function. Since it is irrational to choose an option that is known to be dominated, it follows that it is irrational to use a non-probabilistic credence function. 

An interesting fact about the accuracy argument for probabilism is that it does not depend for its validity on any specific interpretation of the credence function, nor does it depend on the assumption that the ideal credibility function is the function that assigns 1 to the hypothesis that is true and 0 to all hypotheses that are false. Indeed, nothing in the accuracy argument prevents us from designating the ideal credibility function otherwise. Hence, we can easily adapt the argument to a context where the goal is to identify the hypothesis that is best rather than true. In such a context, the ideal function would clearly be one that assigns 1 to the hypothesis that is best and 0 to all other hypotheses. We can then formulate the following version of the accuracy argument:

\begin{quote} P1: The ideal credibility function is the function that assigns 1 to the hypothesis that is best and 0 to all other hypotheses. \end{quote}

\begin{quote} P2: Given any non-probabilistic function, there is a probabilistic function that is guaranteed to have a smaller divergence from the ideal function (given that the divergence measure has certain reasonable properties). \end{quote}

\begin{quote} P3: Given any probabilistic function, there does not exist any function that is guaranteed to have a smaller divergence from the ideal function (given that the divergence measure has certain reasonable properties).  \end{quote}

\begin{quote} P4: If P1-P3, then non-probabilistic credibility functions are irrational. \end{quote}

\begin{quote} C: Non-probabilistic credibility functions are irrational. \end{quote}

P2 and P3 are mathematical theorems (proven by \cite{predd2009}) that hold regardless of what we choose as the ideal function. P1 and P4, on the other hand, are intuitively reasonable general rational principles. The main question that may be raised about the generalized version of the accuracy argument is whether the conditions on the divergence measure are still reasonable when truth is no longer the goal. For example, P2 and P3 require the assumption that the divergence measure belong to the class of Bregman divergences. Is this a reasonable requirement to make? My only response to this question is that I do not see how this assumption (and other necessary mathematical assumptions) are more plausible if truth is the goal than if the goal is to identify the hypothesis that is best in some other sense. So, at least in my eyes, the generalized accuracy argument is at least as plausible as the original argument. In any case, my main goal in this paper is not to give a careful analysis of the accuracy argument. From now I will assume that any credibility function ought to be probabilistic. That is, I will assume that if $p$ is a function that assigns a number between 0 and 1 to each hypothesis $H$ that represents how plausible it is that $H$ is best (in some sense), then $p$ ought to be probabilistic. In the next section, I turn to the main question of the paper: given a probability function $p$ and given a piece of evidence $E$, how should $p$ be updated in light of $E$?

\section{Deriving the updating rules}

Suppose we have a credibility function defined on a hypothesis set \textbf{H} that is probabilistic in the sense of the preceding section. Suppose, also, that we have an evidential measure function $\Ev[E|H]$ defined on the set of evidence and the set of hypotheses under consideration. Note that we are not assuming that $\Ev[E|H]$ is probabilistic (e.g. $\sum_i \Ev[E|H_i]$ need not sum to 1). It is widely accepted that if the goal is to find the true hypothesis in a partition of hypotheses and the evidential measure is the likelihood, i.e. $\Ev[E|H] = p(E|H)$, then any probability function over the hypotheses ought to be updated through Bayesian updating:

\begin{quote} \textbf{Bayesian updating:} $p_E(H) = \frac{p(E|H)p(H)}{\sum_i {p(E|H_i)p(H_i)}}$ \end{quote}

The natural generalization of Bayesian updating is what I have called inferential updating in the introduction. However, it is not clear why the prior probability function and the evidential measure should always be combined in a Bayesian-like manner, regardless of what the evidential measure is and regardless of what the purpose of updating is. Unfortunately, whereas the accuracy argument for probabilism does not make any assumptions about how the credibility function is interpreted, the standard accuracy argument for Bayesian updating \citep{greaveswallace2006} relies on properties that are unique to the likelihood, in particular the fact that the likelihood forms a joint distribution with the prior. Thus, the standard accuracy argument does not generalize to cases where the evidential measure is not the likelihood. Other standard arguments for Bayesian updating have the same limitation (e.g. Dutch book arguments). A different kind of approach is therefore needed.

\cite{bissiri2016} come up with a different approach. They show that provided that the evidential measure is a function of an additive loss function, $L(E, H)$, such that $\Ev[E_1 \& E_2|H] = f(L(E_1, H) + L(E_2, H))$, and given that a few other assumptions are met, then the updating procedure must have the following form, where $c$ is some constant: 

\begin{equation} p_E(H) = \frac{e^{-c*L(E|H)}p(H)}{\sum_i {e^{-c*L(E|H_i)}p(H_i)}} \end{equation}

\cite{bissiri2016} call the above updating procedure ``general Bayesian updating.'' General Bayesian updating traces back to \cite{zhang2006} and been increasingly influential in statistics in recent years.\footnote{See \cite{grunwald2017} for a thorough discussion of general Bayesian updating and related updating rules.} Although \citeapos{bissiri2016} argument for general Bayesian updating is interesting, it has several limitations. One problem is that, as \cite{vassend2019a} argues, the probabilities in (3.1) cannot be interpreted in the standard Bayesian way as plausibilities of truth. But if the probabilities are not standard credibility functions, then the decision theoretic framework assumed by \cite{bissiri2016} would seem to lack justification. The argument also makes certain mathematical assumptions that seem hard to justify from a philosophical point of view. In particular, the authors base their argument in part on the use of statistical divergence measures, and they assume that the divergence belongs to the class of \emph{f}-diverences.\footnote{They also give an alternative derivation that does not make this assumption. However, the alternative derivation makes other suspect assumptions. In particular, it assumes that the normalization procedure is multiplicative, which we'll see later in this paper can be put into question.} This assumption rules out many standard divergence measures, including all Bregman divergences aside from the Kullback-Leibler divergence \citep{amari2009}.\footnote{Recall that Bregman divergences play a crucial role in the accuracy argument for probabilism. The justification for the focus on Bregman divergences is their tight connection to strict propriety (see \cite{predd2009}).} A final limitation of \citeapos{bissiri2016} derivation is that there are many reasonable evidential measures that cannot be written as a function of an additive loss function. Indeed, even the likelihood will only have such a form if the evidence is independent conditional on $H_i$, for all $i$.\footnote{If $p(E_1, E_2 | H) = p(E_1|H)p(E_2|H)$, we can write $p(E_1, E_2 | H) = e^{\log{p(E_1|H)} + \log{p(E_2|H)}}$, i.e. the likelihood is of the form required by \cite{bissiri2016}. But if $p(E_1, E_2 | H) \neq p(E_1|H)p(E_2|H)$, then we cannot write the likelihood in this way.}  Thus, although their argument is interesting, a more general approach that makes less restrictive and more philosophically defensible assumptions is desirable. That is the goal of this section. Later we will see that \citeapos{bissiri2016} updating rule may be derived as a special case.

To start, note that ordinary Bayesian updating can be decomposed into two steps: 

\begin{quote} \textbf{Combination step.} For each $i$, calculate $p^{*}(H_i) =  p(E|H_i)p(H_i)$. 

\textbf{Normalization step.} Transform $p^{*}$ to $p'$ as follows: for each $i$, $p'(H_i) = \frac{p^{*}(H_i)}{p(E)}$.  \end{quote} 

In the first step, the prior plausibility of the hypothesis is combined with the evidential score (i.e. likelihood) of the hypothesis in order to produce an overall judgment of the hypothesis's posterior plausibility. In the second step, the posterior plausibility of all the hypotheses are rescaled in such a way that they jointly obey the probability axioms, i.e. such that all the posterior plausibility scores fall between 0 and 1, inclusive, and jointly sum to 1.

Bayesian updating is a special case of a much broader class of updating rules that decompose into a combination step and a normalization step. The purpose of the remainder of this paper will be to study this class of updating rules. The combination step requires a combination function, $c$, that takes as its input a prior probability, $p(H)$ and a set of evidential scores, $\Ev[E_1 | H]$, $\Ev[E_2 | H, E_1]$, $\Ev[E_3 | H, E_1, E_2]$, etc., and that assigns a total score to $H$, taking into consideration both its prior probability and its performance on the evidence. The normalization step then transforms those scores into probabilities. In other words, on an abstract level, our purpose will be to study updating procedures that decompose in the following way:

\begin{quote} \textbf{Combination step:} For each hypothesis, $H_i$, a set of evidential scores and a prior probability are combined using some combination function $c$ in order to produce an overall posterior score for $H_i$.  \end{quote}
\begin{quote} \textbf{Normalization step:} The posterior scores of all the $H_i$ are transformed using some function $N$ such that they jointly satisfy the probability axioms.  \end{quote}

In the next two subsections the combination step and the normalization step are analyzed in detail. The goal is to show that -- given reasonable assumptions -- the combination function $c$ and the normalization function $N$ both have a very limited set of possible functional forms.

\subsection{The combination step}

Let $e_1$ and $e_2$ represent the evidential scores of a hypothesis $H$ on some evidence, and let $h$ represent $H$'s prior probability; then there are two candidate forms for the combination function that arguably stand out as being particularly plausible:

\begin{quote} \textbf{Additive combination:} $c(e_1, e_2, h) = e_1 + e_2 + h$ \end{quote}
\begin{quote} \textbf{Multiplicative combination:} $c(e_1, e_2, h) = e_1*e_2*h$ \end{quote}

Note that $e_1$ and $e_2$ here may represent either conditional or unconditional evidential scores. For example, $e_1$ may represent  $\Ev[E_1|H]$, i.e. the unconditional evidential score of $H$ on $E_1$, or it may represent $\Ev[E_1|H, E_2]$, i.e. the conditional evidential score of $H$ on $E_1$ given that $E_2$ has already been taken into account. Note, also, that to say that the combination function is additive or multiplicative is not the same as saying that the evidential measure is additive or multiplicative in the sense that $\Ev[E_1, E_2|H] = \Ev[E_1|H] + \Ev[E_2|H] $ or $\Ev[E_1, E_2|H] = \Ev[E_1|H]*\Ev[E_2|H]$. The latter assumptions are much stronger, and amount to assuming that $E_1$ and $E_2$ are independent conditional on $H$ (relative to the evidential measure $\Ev$).

If we make a few reasonable assumptions, we can prove that the combination function must be multiplicative or additive. First of all, suppose we have evidential scores $e_1$ and $e_2$, and a prior probability $h$. Clearly, the order in which we combine the evidential scores and the prior should not matter for the final result we get. That is not to say that the order in which the evidence is received does not matter; it may. For example, if we flip a coin and the outcomes are six heads in a row and then six tails in a row, then the order of the outcomes strongly suggest that the outcomes are probabilistically dependent. Nevertheless, the order in which we evaluate the available pieces of evidence in order to produce an overall judgment should not influence the overall judgment at which we arrive. For that reason, the combination function should be commutative: $c(e_1, e_2) = c(e_2, e_1)$. Furthermore, it clearly should not matter whether we first combine $e_1$ and $e_2$ and then combine the result of that with $e_3$, or whether we combine $e_2$ with $e_3$ and then combine the result with $e_1$, or whether we combine all three pieces of evidence at the same time. In other words, $c$ should be associative: $c(e_1, c(e_2, e_3)) = c(c(e_1, e_2), e_3) = c(e_1, e_2, e_3)$. 

The final reasonable requirement is more quantitative. Clearly, the impact that $e_1$ has on $H$'s overall evidential score, after $e_2$ has already been taken into account, should not depend on the impact that $e_2$ has on $H$. That is not to say that a piece of evidence $E_2$ should not influence the impact that a different piece of evidence $E_1$ has on $H$'s evidential score; it may well, but if it does it should do so through $\Ev[E_1|H, E_2]$. A piece of evidence may influence the evidential impact conferred by another piece of evidence, but the evidential scores themselves should not influence each other. In other words, the requirement is that the impact that, for example, $e_1 = \Ev[E_1|H, E_2]$ makes on $H$'s total evidential score should not depend on the impact that $e_2 = \Ev[E_2|H]$ makes on $H$'s total evidential score, nor vice versa. 

Given that we are willing to suppose that the combination function is twice differentiable, the preceding requirement may be naturally formalized as constraints on the partial derivatives of the combination function. Let $c(x, y)$ be the combination function as a function of variables $x$ and $y$. Then the impact that the evidential score $e_1$ makes on $H$'s total evidential score is plausibly the value of the partial derivative of $c(x, y)$ with respect to $x$, when evaluated at $x = e_1$. If $\frac{\partial{c(x, y)}}{\partial{x}}c(x = e_1,y)$ is a large number, then that means setting $x$ to $e_1$ makes a large difference to $H$'s overall evidential score; if it is 0, then $e_1$ makes no difference. 

The requirement that the impact that $e_1$ makes should not depend on the impact that $e_2$ makes, nor vice versa, for any $e_1$ and $e_2$, may then be formalized in terms of a constraint on the higher-order partial derivatives of $c$, namely that for some constant $k$ the following equation be obeyed:

\[ \frac{\partial^2{c(x, y)}}{\partial{x} \partial{y}} = k \]

The above equation formalizes the idea that the impact that $x$ makes, i.e. $\frac{\partial{c}}{\partial{x}}$, should not depend on the impact that $y$ makes, i.e. $\frac{\partial{c}}{\partial{y}}$, where $x$ and $y$ represent any possible evidential scores. We can now show the following (the derivation is in Appendix A):

\begin{quote} \textbf{Characterization of the combination function.} \emph{Suppose the combination function, $c(x, y)$ satisfies the following requirements}: 

\begin{enumerate}
\item $c$ is commutative. 
\item $c$ is associative. 
\item $c$ is twice differentiable.
\item $c$'s partial derivatives satisfy the following equation, for some number $k$:
\[ \frac{\partial^2{c(x, y)}}{\partial{x} \partial{y}} = k \]
\end{enumerate}
\emph{Then $c$ must have one of the following two forms}: 
\begin{enumerate}
\item If $k = 0$, then $c(x, y) = x+y$.
\item If $k \neq 0$, then $c(x, y) = xy$.
\end{enumerate}

\end{quote}

Hence, it follows that the combination function must be additive or multiplicative. Of course, this conclusion is only as plausible as the assumptions from which it is derived, and some people may be uncomfortable with some of the assumptions that have been made, in particular the condition on the partial derivatives of the combination function. As it happens, it's possible to derive the conclusion from quite different assumptions. Hence, in order to show the robustness of the conclusion, I provide an alternative characterization of the combination function in Appendix F.

\subsection{The normalization step}

After the combination function has produced a posterior plausibility score, the posterior score must be normalized to be a probability. In theory, normalizing a set of numbers means transforming the numbers in such a way that they are all between 0 and 1 and jointly sum to 1, while at the same time retaining as much of their internal structure as possible. In practice, this means that the most extreme numbers in the set may be forced to take the value 0, while the remaining numbers in the set are rescaled by some function, $f$. In other words, normalization in general takes the following functional form:

\begin{equation}   
N(x) = 
     \begin{cases}
       0 &\quad\text{Given that $x$ is sufficiently low}\\
       f(x) &\quad\text{Otherwise} \\
     \end{cases}
\end{equation}

For example, in the normalization step of standard Bayesian updating, $N(x) = f(x)$ (i.e. no non-zero numbers are normalized to 0) and if the set to be normalized is $\{a_1, a_2, \ldots, a_n\}$, then $f(x) = \frac{1}{\sum_i a_i}$. Note that both $N$ and $f$ are relative to the set that is being normalized; hence, if we need to be precise, we should write $N_S$ and $f_S$, where the subscript indicates the set that is being normalized. Nevertheless, I will typically leave off the subscripts in order to avoid clutter. 

Clearly, $f$ should be a one-to-one function. Indeed, except in the case where $x$ and $y$ are both normalized to 0, it should be the case that if $x < y$ then $f(x) < f(y)$. Furthermore, it is clear that the function $f$ ought to commute with the combination function. Suppose we have scores $e_1$, $e_2$, and $h$. Then we should arrive at the same posterior probability regardless of whether we do either of the following: first we combine $h$ and $e_2$, normalize, then combine the normalized result with $e_1$ and normalize again; or we first combine $h$ and $e_1$, normalize, and then combine that normalized result with $e_2$ before normalizing again. In symbols, we require, for all possible scores $x$, $y$, and $z$, that: $f(c(x, f(c(y, z)))) =  f(f(c(x, y), z))$. The justification for this requirement is, again, that the order in which we evaluate our evidence -- which is arbitrary -- should not have an influence on our final judgment. By combining just the preceding two requirements, we can show the following:

\begin{quote} \textbf{Characterization of the normalization procedure.} \emph{Suppose we have a normalization procedure as in (3.2) that satisfies the following requirements}: 

\begin{enumerate}
\item $f$ commutes with the combination function $c$. For all $x$, $y$, and $x$:  $f(c(x, f(c(y, z)))) =  f(f(c(x, y), z))$.
\item $f$ is one-to-one: for all $x$ and $y$, $f(x) = f(y)$ if and only if $x = y$.
\end{enumerate}
\emph{Then the normalization process must have one of the following forms, for some constant $k$ that depends on the set, $S$, of numbers being normalized}: 
\begin{enumerate}
\item If the combination function is multiplicative, then, for all $x$ in $S$, $f(x) = k*x$.  
\item If the combination function is additive, then, for all $x$ in $S$,  $f(x) = x + k$.  
\end{enumerate}

\end{quote} 

The proof, which again is straightforward, is in Appendix B.

\subsection{Characterizations of inferential and predictive updating}

The results so far show that any updating procedure needs to have either: (1) A multiplicative combination step and a multiplicative normalization step, or (2) an additive combination step and an additive normalization step. Call an updating procedure that satisfies either (1) or (2) a \textbf{legitimate} updating procedure.\footnote{Note that not every updating rule that has been suggested in the literature is legitimate in this sense of the word. For example,  \cite{douven2017} consider a rule according to which $p_E(H) = c*(p(H)*p(E|H) + f(E, H))$ where $c$ is a normalization constant and $f(E, H)$ is a ``bonus'' assigned to $H$ in case $H$ is the best explanation of $E$. This updating rule is not legitimate because it is neither purely additive nor purely multiplicative. On the other hand, the class of rules considered in \cite{douven2016} are legitimate. }

To characterize inferential updating we now introduce the following principle:

\begin{quote} \textbf{Regularity:} No hypothesis is ever conclusively ruled out by any evidence unless the evidence logically refutes the hypothesis, i.e. the posterior probability of any hypothesis is always greater than 0. \end{quote}

We can then show the following (see Appendix C): 

\begin{quote} \textbf{Characterization of inferential updating.} The only legitimate updating procedure that satisfies Regularity is inferential updating. I.e., given evidential measure $\Ev$ and prior probability function $p$, update $p$ to the posterior $p_E$ by way of the following formula:

\[ 
p_E(H) = \frac{\Ev{[E | H]}p(H)}{\sum_i \Ev{[E | H_i]}p(H_i)} 
\]

\end{quote}

Inferential updating satisfies Regularity; it will never result in any hypothesis having a posterior probability of 0. On the other hand, in Appendix C, I show that an updating procedure that uses an additive combination function and an additive normalization function must violate Regularity; most of the time, any such updating rule must assign a posterior probability of 0 to some hypotheses. But this does not mean that such an updating rule should never be used. As we will see in the next section, sometimes we may want to be able to exclude certain hypotheses from consideration---i.e., assign them a posterior probability of 0.

Nevertheless, we do not want to exclude more hypotheses than is warranted by the data. The updating procedure ought to be conservative and exclude as few hypotheses as possible at every step. In other words, any updating procedure that violates Regularity should plausibly still satisfy the following principle:

\begin{quote} \textbf{Conservativeness:} The updating procedure assigns a posterior probability of 0 to as few hypotheses as possible, given the combination function, the normalization procedure, and the evidence available. \end{quote}

We are now in a position to characterize predictive updating:

\begin{quote} \textbf{Characterization of predictive updating.} The only legitimate updating procedure that violates Regularity, but satisfies Conservativeness, is predictive updating. I.e., given evidential measure $\Ev$ and prior probability function $p$, update $p$ to the posterior $p_E$ by way of the following procedure:

\begin{quote} \textbf{Step 1.} For each $i$, calculate $q(H_i) =  p(H_i) + \Ev[E|H_i]$. 

\textbf{Step 2.} Transform $q$ to $p_E$ as follows: for each $i$, $p_E(H_i) = 0$ or $p_E(H_i) =  q(H_i) + d$, where $d$ is the unique number such that $d$ is minimal and, for all $i$, $p_E(H_i) \ge 0$ and $\sum_i {p_E(H_i) = 1}$.  \end{quote} 

 \end{quote}

\section{Discussion of inferential and predictive updating}

\subsection{The difference between inferential updating and predictive updating}

Inferential updating and predictive updating differ in that the former updating rule obeys Regularity while the latter rule does not. Is Regularity a reasonable constraint? In some contexts it is, but in others it is not. Suppose our main priority is to identify the hypothesis that is true or (if none of the hypotheses is true) the hypothesis that is closest to the truth according to some appropriate measure of closeness to the truth. Given this goal, it is reasonable to be risk-averse and open-minded: we do not want to rule out any hypothesis as potentially being the hypothesis that is true. Even if a lot of evidence strongly suggests that a hypothesis is false, there is always the possibility that the evidence is unrepresentative or misleading. And so Regularity is a reasonable constraint in this context.

However, suppose we do not care about which of our hypotheses is true or closest to the truth; our goal is not inferential, but predictive. We wish to find, as efficiently as possible, the subset of hypotheses that can be expected to be as predictively accurate as possible. In this context, there is no theoretical justification for requiring that the updating rule obey Regularity; on the contrary, there are good reasons for why we might want an updating rule that violates Regularity. In particular, suppose the posterior distribution will be used in order to make a weighted probabilistic prediction, i.e. the goal is for $p(D|H_i)p_E(H_i)$ to be as accurate on future data $D$ as possible. In that case, it would seem inadvisable to assign positive probability to any hypothesis that has shown itself to be very predictively inaccurate, since the predictions made by such a hypothesis would likely throw off the weighted prediction. On the other hand, we do not want to go to the opposite extreme and base the prediction on the single hypothesis that has performed best on the evidence, as that is liable to lead to overfitting \citep{forstersober1994}. Predictive updating enables one to set the probabilities of predictively inaccurate hypotheses to 0 in a principled (and conservative) way. 

Let's consider a specific example. When the hypotheses under considerations make probabilistic predictions and the goal is maximal predictive accuracy, it is natural to use a strictly proper scoring rule as the measure of evidential favoring \citep{gneiting2007}. For various reasons, the most popular scoring rule in applied research is probably the Continuous Ranked Probability Score (CRPS). Suppose we have a set of competing statistical models $M_1$, $M_2$, etc., and for each model, let $p_{M_i}$ be the marginal (cumulative) probability forecast distribution corresponding to $M_i$. Suppose, moreover, that $p_{M_i}$ has finite first moment, that $X$, $X_1$ and $X_2$ are independent and identically distributed random variables that follow the distribution of $p_{M_i}$, and that $x$ is the actual observed outcome. Then the CRPS can be written in the following way (where the expectations are taken relative to $p_{M_i}$): 

\begin{equation} \mathrm{CRPS}(p_{M_i}, x) = \E|X - x| - \frac{1}{2}\E|X_1 - X_2| \end{equation} 

As (4.1) makes clear, CRPS is a statistical generalization of absolute error. As \cite{gneiting2007} point out, a significant benefit of the CRPS is that it is easily interpretable, since the outputs of (4.1) can be reported in the same units as the measurements. For example, suppose the measurements are in terms of meters. Then the CRPS score of a model on an observation will be a representation of how many meters inaccurate the model's predictions are of that observation, on average (since the prediction is a probability distribution rather than a single number, the average is needed).

If we let $\Ev[x|p_{M_i}] = a*\mathrm{CRPS}(p_{M_i}, x)$, where $a$ is some constant, and assign prior probabilities to all the models, then predictive updating can be used to assign posterior probabilities to all the models.\footnote{If the models contain parameters, then the probability distributions over those parameters may be updated using either inferential or predictive updating.} Importantly, given sufficient evidence (and depending on how the constant $a$ is chosen) many of the models will receive a posterior probability of 0. These posterior probabilities can then be used for model selection or for making a weighted prediction using all the models. Of course, it is an empirical question whether predictive updating is better (for predictive purposes) than inferential updating (including standard Bayesian updating). An empirical evaluating of predictive updating will have to wait for a different occasion, however. In this section I have simply tried to suggest one concrete way in which predictive updating may be implemented.

\subsection{The relationship between inferential updating and other updating procedures}

As was already mentioned in the introduction to the paper, standard Bayesian updating is clearly a special case of inferential updating: more precisely, we get Bayesian updating if and only if $\Ev[E|H] \propto p(E|H)$, i.e. if and only if the evidential measure is proportional to the likelihood. What \cite{vassend2019b} calls ``quasi-Bayesian updating'' is also a special case of inferential updating; indeed, quasi-Baysian updating is simply inferential updating with an evidential measure that has been suitably calibrated to a verisimilitude measure. Similarly, \citeapos{douven2016} IBE-based updating rule is also clearly a kind of inferential updating. 

Perhaps more interestingly, \citeapos{bissiri2016} general Bayesian updating is also a special case of inferential updating. More precisely, we have: 

\begin{quote}
\textbf{General Bayesian updating is a special case of inferential updating.} \emph{Suppose the evidential measure $\Ev$ is a strictly decreasing function $f$ of some loss function, $L(E,H)$, such that for all $E_1$ and $E_2$, $\Ev$ satisfies the following conditions}: 

\begin{enumerate}
\item $\Ev[E_1|H, E_2] = \Ev[E_2|H] = f(L(E_1, H))$.
\item $\Ev[E_1, E_2 | H] = f(L(E_1, H) + L(E_2, H))$ . 
\end{enumerate}

\emph{Then inferential updating has the following form}:

\[ p(H|E) = \frac{e^{-c*L(E, H)}p(H)}{\sum_i {e^{-c*L(E, H_i)}p(H_i)}} \]

\emph{For some constant $c$}.

\end{quote}

A sketch of the proof, which is straightforward, is given in Appendix E. Although general Bayesian updating is a special case of inferential updating, the reverse is not the case because -- as was previously mentioned -- many reasonable evidential measures cannot be written as a function of an additive loss function. Suppose, for example, that the hypotheses under consideration are real-valued functions, $f_i$ and that the evidential measure is of the form $\Ev[(x_1, y_1), (x_2, y_2), \ldots, (x_n, y_n)|f_i] = \mathrm{Minimum}(|y_1 - f_i(x_1)|, |y_1 - f_i(x_1)|, \ldots, |y_1 - f_i(x_1)|)$. It is clear in this case that the evidential measure cannot be written as a function of an additive loss function, simply because the Minimum operator is not additive. 

A diagram depicting the relationship between inferential updating, predictive updating, and various updating rules that have been suggested in the literature is given in Figure 1.

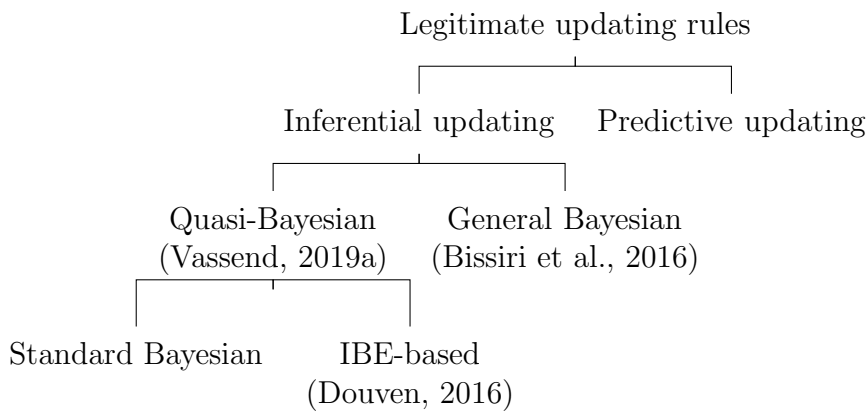
\begin{figure}[!h]
\begin{forest}
  for tree={
    align=center,
    edge path={
      \noexpand\path [draw, \forestoption{edge}] (!u.parent anchor) -- +(0,-15pt) -| (.child anchor)\forestoption{edge label};
    },
    if n=2{
      if ={equal(n_children("!u"),3)}{calign with current}{}}{},
  }
  [Legitimate updating rules
    [Inferential updating
    [Quasi-Bayesian \\ \citep{vassend2019b}
    [Standard Bayesian][IBE-based \\ \citep{douven2016}]
    ]     [General Bayesian \\ \citep{bissiri2016}] ]
    [Predictive updating
    ]
  ]
\end{forest}
\caption{Overview of various updating rules}
\end{figure}

\section{Conclusion}

The primary purpose of this paper has been to justify a set of very general synchronic and diachronic inductive norms. The resulting normative framework can be put to both philosophical and scientific use. In philosophy of science, a standard way of analyzing scientific methodology is by seeing whether the methodology makes sense from a Bayesian perspective. For example, in this way, \cite{sober2015} analyzes parsimony inference,\footnote{Sober uses a likelihoodist approach, which is Bayesianism without the priors.} \cite{dawid2015} analyze no-alternatives arguments in physics, \cite{schupbach2018} analyzes robustness analysis, and \cite{myrvold2016} evaluates the epistemic value of unification. Since the preceding analyses take place in a Bayesian framework, they inherit the limitations and assumptions of Bayesianism. In the broader normative framework developed in this paper, it's possible to check whether the analyses still hold up when those assumptions are lifted. For example, \cite{myrvold2016} shows that more unifying hypotheses will be more confirmed by evidence than less unifying hypotheses, other things being equal. Since his analysis is Bayesian, he implicitly uses the likelihood as his measure of evidential favoring. A natural question to ask is whether his result still holds if the likelihood is replaced with an arbitrary measure of evidential favoring. The perhaps surprising answer is yes, although a proper demonstration of this fact must be reserved for a different time.

The normative framework developed in this paper can also be used for scientific inference. Indeed, implicitly it already has been---as shown in Section 4.2, the general Bayesian updating rule suggested by \cite{bissiri2016} is a special case of inferential updating, and general Bayesian updating is gaining in popularity in the statistical community. But inferential updating is more general than general Bayesian updating, and allows for the use of evidential measures that cannot be represented in \citeapos{bissiri2016} framework. One example is the phylogenetic parsimony measure discussed by \cite{vassend2019b}. Predictive updating can also be applied in scientific inference problems, for example through the use of strictly proper scoring rules as suggested in Section 4.1. Of course, it is ultimately an empirical question whether predictive updating performs better than inferential updating. An answer to this question must wait until later; in this paper, my goal has been to provide a general normative framework for inductive inference that is as flexible as possible while obeying basic theoretical desiderata.

\appendix

\section{Characterization of the combination function}

The goal of this section is to show the characterization of the combination function in Section 3.1. There are two cases to consider: $k=0$ and $k \neq 0$. Since the two cases are very similar, I will only consider the case where $k \neq 0$. So suppose that for some non-zero $k$, we have:

\begin{equation} \frac{\partial^2{c(x, y)}}{\partial{x} \partial{y}} = k \end{equation}

Taking the antiderivative with respect to $x$, it follows that: 

\begin{equation}  \frac{\partial{c(x, y)}}{\partial{y}} = kx + C(y) + D \end{equation}

Where $C(y)$ is a function of $y$, but not $x$, and $D$ is some real number. Taking the antiderivative of (A.2) with respect to $y$, we get:

\begin{equation}  c(x, y) = kxy + \int C(y) dy + Dy + G(x) + F \end{equation}

Where $G$ is a function of $x$ and $F$ is some real number. Moreover, exchanging the labels $x$ and $y$ in (A.3) gives us:

\begin{equation}  c(y, x) = kyx + \int C(x) dx + Dx + G(y) + F \end{equation}

But since $c(x, y) = c(y, x)$, (A.3) and (A.4) must be equal, which means that $kxy + \int C(y) dy + Dy + G(x) + F = kxy + \int C(x) dx + Dx + G(y) + F$, and hence $\int C(y) dy + Dy + G(x) = \int C(x) dx + Dx + G(y)$. Rearranging, we get: 

\begin{equation} G(x) = \int C(x) dx + Dx + G(y) - \int C(y) dy - Dy  \end{equation}

 But since $G(x)$ does not depend on $y$, the only way for (A.5) to be true is if $G(y) - \int C(y) dy - Dy$ is equal to some constant number, $c$. Hence, $\int C(y) dy + Dy = G(y) - c$ . Plugging this back into (A.3) (and absorbing the constant $c$ into $F$), we get: 

\begin{equation}  c(x, y) = kxy + G(x) + G(y) + F \end{equation}

Without loss of generality, we may assume that $G(0) = 0$, because if $G(0) = A$ for some non-zero $A$, then we can just put $G'(x) = G(x) - A$ and $F' = F + 2A$, and we get: $c(x, y) = kxy + G'(x) + G'(y) + F'$, with $G'(0) = 0$ (i.e. we simply absorb the constant $A$ into $F'$).

Now the fact that $c$ is associative and commutative means that $c(c(x, y), z)) = c(c(y, z), x)$, and hence (A.6) implies that, for all $x$, $y$, and $z$:

\begin{equation}
\begin{gathered}   
k(kxy + G(x) + G(y) + F)z + G(kxy + G(x) + G(y) + F) + G(z) + F \\
= k(kyz + G(y) + G(z) + F)x + G(kyz + G(y) + G(z) + F) + G(x) +F  
\end{gathered}   
\end{equation}
 
 Simplifying, we have:
 
\begin{equation}
\begin{gathered}     
[G(x) + G(y) + F]kz + G[kxy + G(x) + G(y) + F] + G(z) \\
= G(y)kx + G(z)kx + Fkx + G[kyz + G(y) + G(z) + F] + G(x) 
\end{gathered}   
\end{equation}

Note that because $c$ is twice differentiable, so is $G$. Taking the derivative of each side of (A.8) with respect to $z$ gives:

\begin{equation}  [G(x) + G(y) + F]k +  \frac{\partial{G(z)}}{\partial{z}} = \frac{\partial{G(z)}}{\partial{z}}kx + G'[kyz + G(y) + G(z) + F]*\frac{\partial{G(z)}}{\partial{z}} \end{equation}

Next, taking the derivative of each side of (A.9) with respect to $x$ gives:

\begin{equation}  \frac{\partial{G(x)}}{\partial{x}}k = \frac{\partial{G(z)}}{\partial{z}}k \end{equation}

Hence, since $k \neq 0$, it follows that $\frac{\partial{G(x)}}{\partial{x}} = \frac{\partial{G(z)}}{\partial{z}}$. But since $G(x)$ does not depend on $z$ and $G(z)$ does not depend on x, this means that $\frac{\partial{G(x)}}{\partial{x}}$ must be a constant number, i.e. $\frac{\partial{G(x)}}{\partial{x}} = a$ for some constant $a$. Since we are assuming that $G(0) = 0$, it follows that $G(x) = ax$. Next, the fact that $c(x, y, z) = c(c(x, y), z)$ implies:
 
 \begin{equation}  kxyz + ax + ay + az + F = k(kxy + ax + ay + F)z + a(kxy + ax + ay + F) + az + F  \end{equation}
 
Comparing the terms that contain $xyz$,\footnote{Which we can do, as before, by successively differentiating with respect to $x$, $y$, and $z$. This proof method is sometimes called ``equating coefficients'' \citep[p. 169]{tanton2005}.} we see that $k=1$, and hence:
 
  \begin{equation}  ax + ay =  axz + ayz + Fz + axy + a^2x + a^2y + Fa  \end{equation}
  
Comparing the terms that contain $z$, we see that $a(x + y) + F = 0$ for all $x$ and $y$. The only way this can be true is if $a = F = 0$. Hence we have, finally, that $c(x, y) = xy$.

 \section{Characterization of the normalization step}
 
 The goal of this section is to show the characterization of the normalization step in Section 3.2. Let $\{a_i\}$ be an arbitrary set of $n$ numbers, $S_1$, with normalization function $f_{S_1}$. Consider the set $S_2 = \{\frac{1}{a_i}\}$ and the set $S_3 = \{1_i\}$, which consists of $n$ copies of 1. Then condition (1) implies that, for all $i$, $f(c(f(c(\frac{1}{a_i}, a_i)), 1)) = f(c(\frac{1}{a_i}, f(c(a_i, 1))))$, where the various $f$'s are relative to the relevant sets. For example, in $f(c(\frac{1}{a_i}, a_i))$, $f$ is a rescaling function defined on the set $\{c(\frac{1}{a_i}, a_i)\}$ . Note that we are abusing notation here: strictly speaking the various $f$'s are not the same function, since they are defined over different sets. However, to avoid needless clutter, I use $f$ without subscripts.

According to the characterization of the combination function, the combination function is either multiplicative or additive. Since the derivations are very similar, I will only show that the normalization function must be multiplicative given that the combination function is multiplicative. So suppose that the combination function is $c(a, b) = ab$. Then we get: $f(f(\frac{1}{a_i}*a_i)*1) = f(\frac{1}{a_i}* f(a_i *1))$. Thus, we have: $f(f(1)) = f(\frac{1}{a_i}* f(a_i))$, i.e. $f(\frac{1}{a_i}* f(a_i))$ is a constant. But since, $f$ is one-to-one, that means $\frac{1}{a_i}* f(a_i)$ must also be a constant. That is, there exists a constant $k$ such that, for all $a_i$ in $S$, $\frac{1}{a_i}* f(a_i) = k$. Hence $f(a_i) = k*a_i$ for all $a_i$. Since $S$ was an arbitrary set, it follows that in general the normalization procedure must be multiplicative given that the combination function is multiplicative.

\section{Characterization of inferential updating}

The goal in this section is to show that the only legitimate updating rule that satisfies Regularity is inferential updating. According to the results in sections 3.1 and 3.2, any legitimate updating rule must either have (1) a multiplicative combination step and a multiplicative normalization step, or (2) an additive combination step and an additive normalization step. It is easy to show that it is possible for an updating rule that satisfies (1) to satisfy Regularity, and that -- indeed -- the resulting updating rule is inferential updating. In order to show that inferential updating is the only updating rule that satisfies Regularity, it suffices to show that there is no updating rule satisfying (2) that also satisfies Regularity. 

Suppose, for the sake of contradiction, that there is some updating rule that satisfies both (2) and Regularity. In order for Regularity to be obeyed, it has to be the case that given any set of non-zero prior probabilities over a set of hypotheses, $h_1, h_2, \ldots, h_n$, and given any set of evidential scores for the hypotheses, $e_1, e_2, \ldots, e_n$, the posteriors are also all non-zero. Thus, if $N$ is the normalization function, then the following must be true for all $h_i$:
 
 \begin{equation} N(e_i + h_i) > 0 \end{equation}
 
 Since the normalization function is assumed to satisfy (2), C.1 implies that the following is true for all $i$, where $d$ is an additive normalization constant: 
 
  \begin{equation} e_i + h_i + d > 0 \end{equation}
  
  Since the posterior probabilities must sum to 1, we also have:
  
    \begin{equation} \sum_i {(e_i + h_i + d)}  = 1 \end{equation}
    
 And therefore, $d = -\frac{1}{n}\sum e_i$. And so we have, for all $h_i$:
 
 \begin{equation} e_i + h_i -\frac{1}{n}\sum e_i   > 0 \end{equation}
 
 But it's obvious that (D.4) will not in general be true. For example, suppose $e_1$ is the smallest $e_i$. Then $r = e_i - \frac{1}{n}\sum e_i < 0$. Now suppose it's also the case that $h_1 < -r$. Then we have:

 \begin{equation} e_1 + h_1 -\frac{1}{n}\sum e_i   = r + h_1 < 0 \end{equation}
 
 Consequently, additive combination and additive normalization jointly violate Regularity. So there can be no updating procedure that satisfies both (2) and Regularity.

\section{Characterization of predictive updating}

The goal in this section is to show that the only legitimate updating rule that violates Regularity but satisfies Conservativeness is predictive updating. It is clear that any updating rule that satisfies Conservativeness but violates Regularity must be additive. This is because any multiplicative updating rule that satisfies Conservativeness clearly also satisfies Regularity. 

So suppose the updating rule is additive and satisfies Conservativeness. Then the goal is to show that the updating rule must be equivalent to predictive updating. Since the rule is additive, it must have the following form, where $p_E$ is the posterior probability distribution, $H_i$ is a hypothesis, $h_i$ is the prior probability of the hypothesis, $e_i$ is the evidential score of the hypothesis, and $d$ is a normalization constant:

\begin{equation}   
p_E(H) = 
     \begin{cases}
       0 &\quad\text{Given that $x$ is sufficiently low}\\
       h_i + e_i + d &\quad\text{Otherwise} \\
     \end{cases}
\end{equation}

If the updating rule is conservative, then as few hypotheses as possible should be assigned a posterior probability of 0. It remains to show that this uniquely happens when $d$ is minimal. Suppose there are $n$ hypotheses. Without loss of generality, suppose the hypotheses are ordered such that $ 0 \ge p_E(H_1) \ge p_E(H_2) \ge \ldots \ge p_E(H_n)$. Then there is some index $m$ such that $p_E(H_i) = 0$ for $i \le m$ and $p_E(H_i) > 0$ for $i > m$. Note that the updating procedure is conservative if and only if $m$ is minimal because $m$ is minimal if and only if a minimal number of hypotheses have a posterior probability of 0. In order for the posterior probabilities to be probabilistic, we must have: 

\begin{equation}  
  \sum_i p_E (H_i) =   \sum_{i > m}(h_i + e_i) + (n-m)d = 1
\end{equation}

Now suppose we have a different updating rule resulting in some posterior $p'$ that is \emph{not} conservative: i.e. there an index $m' > m$ such that  $p^{'}_E(H_i) = 0$ for $i \le m'$ and $p^{'}_E(H_i) > 0$ for $i > m'$. Then $p'$ must satisfy the following constraint for some normalization constant $d'$:

\begin{equation}  
     \sum_{i > m'}(h_i + e_i) + (n-m')d' = 1
\end{equation}

Comparing D.2 and D.3 and remembering that $m' > m$, we see that:

\begin{equation}  
    0 < \sum^{m'}_{i=m}(h_i + e_i) = (n-m')d' - (n-m)d
\end{equation}

And hence, 

\begin{equation}  
     d < \frac{n-m'}{n-m}d' < d'
\end{equation}

Hence, $d < d'$. What the above proof shows is that any conservative updating rule has a smaller additive normalization constant than any non-conservative updating rule. To finish the proof, we show that there is just one conservative updating rule. Here we can use D.4 again. If both updating rules are conservative, then we have $m = m'$, and hence -- making the necessary amendments in D.4, we have: 

\begin{equation}  
    0 = \sum^{m'}_{i=m}(h_i + e_i) = (n-m)d' - (n-m)d
\end{equation}

Hence it follows that $d' = d$. But then the two updating rules are equivalent. Hence, there is only one conservative updating rule, namely the one that uses a minimal additive normalization constant. This is predictive updating.  
 
 \section{General Bayesian updating is a special case of inferential updating}

The goal in this section is to show that \citeapos{bissiri2016} general Bayesian updating is a special case of inferential updating. For some normalization constant $k$, we have:

\begin{equation} p(H|E_1, E_2) = k*\Ev[E_1|H, E_2]\Ev[E_2|H]p(H) = k*f(L(E_1, H))f(L(E_2, H))p(H) \end{equation}

But we also have:

\begin{equation} p(H|E_1, E_2) = k*\Ev[E_1, E_2|H]p(H) = k*f(L(E_1, H) + L(E_2, H))p(H) \end{equation}

Comparing C.1 and C.2, we see that $f$ obeys the following functional equation for all $x$ and $y$: $f(x)f(y) = f(x+y)$. Let $g(x) = \log{f(x)}$. Then $g(x + y) = g(x) + g(y)$, which is the well known Cauchy equation whose solution is $g(x) = -cx$, for some positive constant $c$ \citep[p. 31]{aczel2006} (since $f$, and therefore $g$, is strictly decreasing). Consequently $f(x) = e^{-cx}$, and hence $p(H|E) = k*e^{-c*L(E, H)}p(H)$, which is \citeapos{bissiri2016} general Bayesian updating rule.

 \section{An alternative characterization of the combination step}
 
In both everyday and scientific contexts, it's common to think of evidence algebraically: multiple lines of evidence combine in order provide stronger evidence; some evidence favors a hypothesis, while other evidence goes against it; a piece of evidence here can cancel out a piece of evidence there; and some purported evidence has no effect at all. In other words, evidential favoring has all the hallmarks of a mathematical group. Now, suppose -- as we have been doing up to now -- that we use real numbers to represent evidential scores. Then the set of all possible evidential scores, $G$, together with the combination function plausibly form a mathematical group. Indeed, they plausibly form an \emph{Archimedean} group, because intuitively there is no maximal evidential score. That is, if we use $\bullet$ to denote the combination function, i.e. $e_1\bullet e_2 = c(e_1, e_2)$, then it is plausible that $(G, \bullet)$ satisfies the following axioms:

\begin{enumerate}
  \item \textbf{Closure.} For all possible evidential scores $e_1$ and $e_2$, $e_1 \bullet e_2$ is also a possible evidential score.
    \item \textbf{Associativity.} For all possible evidential scores $e_1$, $e_2$ and $e_3$, $(e_1 \bullet e_2) \bullet e_3 = e_1 \bullet (e_2 \bullet e_3)$.
       \item \textbf{Identity.} There exists a possible evidential score $i$ such that for all $e$, $i \bullet e = e \bullet i = e$. I.e., there exists a real number that represents evidence that has no effect (either favorable or unfavorable). 
       \item \textbf{Inverse.} For each possible evidential score  $e$, there exists a possible evidential score $e'$ such that $e \bullet e' = e' \bullet e = i$. I.e. every evidential score could potentially (in principle) be cancelled out by other countervailing evidence.\footnote{A referee points out that this is a bit of an idealization, since a piece of evidence and a defeater of that evidence will not typically cancel each other out precisely.}
       \item \textbf{Commutativity.} For all possible evidential scores $e_1$ and $e_2$, $e_1 \bullet e_2 = e_2 \bullet e_1$. I.e. the order in which the evidence is considered is irrelevant.
         \item \textbf{Archimedean property.} For all possible evidential scores $e_1$ and $e_2$, there exists an integer $n$ such that $e_1 < e_2 \bullet e_2 \ldots \bullet e_2 ($n$ times)$.
  
  \end{enumerate}
  
 Suppose, in addition, that the set of evidential scores is totally ordered: for all evidential scores $e_1$ and $e_2$, either $e_1 > e_2$ or $e_1 \le e_2$.\footnote{A referee rightly points out that this assumption is also idealized.} Then we can use the following important result from group theory (see \cite[p. 33]{kopytov1996}, for a proof):
  
  \begin{quote}
  
  \textbf{H\"{o}lder's theorem.} Every Archimedean totally ordered group is order-isomorphic to a subgroup of the additive group of real numbers with the natural order.
  
    \end{quote}
    
    The fact that $(G, \bullet)$ is order-isomorphic to a subgroup of the additive group of real numbers with the natural order means there exists some subgroup, $(S, +)$ of the real numbers and a one-to-one function, $g$, from $(G, \bullet)$ to $(S, +)$ that obeys the following equation for all $e_1$ and $e_2$ in $G$: $g(e_1 \bullet e_2) = g(e_1) + g(e_2)$. Since $g$ is one-to-one, it has an inverse, $f$. Hence, for all $e_1$ and $e_2$ in $G$, we can write: $e_1 \bullet e_2 = f(g(e_1) + g(e_2))$. 
    
In the main text, I showed that the normalization procedure must be either additive or multiplicative, given that the combination function is either multiplicative or additive. But, arguably, it is not unreasonable to simply assume that the normalization must be either multiplicative or additive. Indeed, all updating rules that have been proposed in the literature have implicitly relied on a normalization procedure that is either multiplicative or additive. In particular, the normalization procedure implicit in both standard Bayesian updating and Jeffrey updating \citep{jeffrey1983} is multiplicative, and the normalization procedure implicit in \citeapos{leitgeb2010} alternative to Jeffrey updating is additive.

Finally, it is reasonable to assume -- as we did in the main text -- that the normalization procedure commutes with the combination function in the sense that, for all $a$, $b$, and $c$, we have: $N(a \bullet N(b)) = N(N(a) \bullet b) = N(a \bullet b)$. We can now give the following characterization of the combination function:

\begin{quote} \textbf{Alternative characterization of the combination function.} \emph{Suppose the combination function, $c(x, y)$ satisfies the following requirements}: 

\begin{enumerate}
\item The set of all evidential scores, $G$, and the combination function $c(x, y) = x \bullet y$ together form a totally ordered Archimedean group.
\item The combination function commutes with the normalization function $N$ in the sense that, for all $a$, $b$, and $c$: $N(a \bullet N(b)) = N(N(a) \bullet b) = N(a \bullet b)$.
\end{enumerate}
\emph{Then $c$ must have one of the following two forms}: 
\begin{enumerate}
\item If the normalization function is additive, then $c(x, y) = x+y$.
\item If the normalization function is multiplicative, then $c(x, y) = xy$.
\end{enumerate}

\end{quote}

\emph{Proof.} The fact that the combination function commutes with the normalization function implies that, for every $e$ with inverse $e^{-1}$:
    
    \begin{equation} N(e \bullet e^{-1}) = N(N(e) \bullet e^{-1}) = N(f(g(N(e)) + g(e^{-1}))) \end{equation}
    
 Therefore, for all $e$, $N(f(g(N(e)) + g(e^{-1}))) = N(i)$, where $i$ is the identity element of the group. Since $N$ is one-to-one, this means that $f(g(N(e)) + g(e^{-1})) = k$, for some constant $k$ that does not depend on $e$. Furthermore, since $f$ is one-to-one, this in turn implies that $g(N(e)) + g(e^{-1}) = k'$, for some constant $k'$ that does not depend on $e$. For the same reason, (F.1) also implies that $g(e) + g(e^{-1}) = k''$, for some constant $k''$ that does not depend on $e$. Hence we have, finally, that $g(N(e)) - g(e) = K$, where $K = k' - k''$. Hence, $g(N(e)) = g(e) + K$. 
   
   If the normalization procedure is multiplicative, then for some normalization constant $a$, we have $g(ae) = g(e) + K$. Note that $a$ depends on the set to which $e$ belongs. If $\{e_i\}$ is the set, then
   
   \begin{equation} a = \frac{1}{\sum e_i} \end{equation}
   
    Hence, depending on the other members of the set to which $e$ belongs, $a$ can be any number in the half-open interval $(0, \frac{1}{e})$. Thus we have, for all $e$ and all $a$ in $(0, \frac{1}{e})$, that $g(ae) = g(e) + K$, where $K$ is a constant that may depend on $a$, but does not depend on $e$.
    
    Similarly, we have---for some normalization constant $b$---that $g(bae) = g(ae) + K' = g(e) + K''$. Here, $b$ can be any number in the range $(0, \frac{1}{ae})$, or in other words in $(0, \infty)$. But if we let $y = ab$ and $x = e$, then the preceding means that for all $x$ and $y$ in $(0, \infty)$ we have: 
    
    \begin{equation} g(yx) = g(x) + K'' \end{equation}
    
    Where $K''$ depends on $y$, but not on $x$. Interchanging the role of $y$ and $x$, we also have: 
    
        \begin{equation} g(xy) = g(y) + K''' \end{equation}
        
        Where $K'''$ depends on $x$, but not on $y$. Comparing the above equations, we see that $g(x) + K'' = g(y) + K'''$. This implies the following:
        
           \begin{equation} g(xy) = g(x) + g(y) + C \end{equation}
           
           Where $C$ is a constant that depends on neither $x$ nor $y$. Now note that $f(2g(i)) = i \bullet i = i = f(g(i)$. Since $f$ is one-to-one, this implies that $g(i) = 0$. Next, (F.5) implies that $g(i) = g(1*i) = g(1) + g(i) + C$. Thus $g(1) = -C$. Using (F.5) again, we have $g(1) = g(i*\frac{1}{i}) = g(i) + g(\frac{1}{i}) = g(\frac{1}{i})$. But since $g$ is one-to-one, this implies that $\frac{1}{i} = 1$, so that $i=1$. Hence $-C = g(1) = g(i) = 0$, so $C = 0$. Finally, then, we have, for all $x > 0$ and $y >0$: 
           
             \begin{equation} g(xy) = g(x) + g(y) \end{equation}

   Now put $r(x) = g(e^{x})$. Then (F.6) becomes, for all real $x$ and $y$:
   
    \begin{equation} r(x + y) = r(x) + r(y) \end{equation}
    
    This is the Cauchy functional equation, whose only solution is $r(x) = cx$, for an arbitrary constant $c$ \citep[p. 31]{aczel2006}. Hence, $g(x) = r(\log x) = \log x^c$. Since $f$ is the inverse of $g$, we have that $f(x) = e^{x^{\frac{1}{c}}}$. Finally, then, we have:

                \begin{equation} x \bullet y = f(g(x) + g(y)) = e^{(\log(x^c) + \log(y^c))^{\frac{1}{c}}} = e^{(c*\log(xy))^{\frac{1}{c}}} = xy \end{equation}
 
I.e. the combination function is multiplicative, $c(x, y) = xy$.
 

\begin{thebibliography}{}

\bibitem[\protect\citeauthoryear{Acz{\'e}l}{Acz{\'e}l}{2006}]{aczel2006}
Acz{\'e}l, J. (2006).
\newblock {\em Lectures on Functional Equations and Their Applications}.
\newblock Dover Books on Mathematics. Dover Publications.

\bibitem[\protect\citeauthoryear{Amari}{Amari}{2009}]{amari2009}
Amari, S.-I. (2009).
\newblock {\text{alpha}-Divergence is Unique, Belonging to Both
  \textit{f}-Divergence and Bregman Divergence Classes}.
\newblock {\em IEEE Transactions on Information Theory\/}~{\em 55\/}(11), 4925
  -- 4931.

\bibitem[\protect\citeauthoryear{Bernardo and Smith}{Bernardo and
  Smith}{1994}]{bernardosmith1994}
Bernardo, J.~M. and A.~F.~M. Smith (1994).
\newblock {\em Bayesian Theory}.
\newblock Wiley, New York, NY.

\bibitem[\protect\citeauthoryear{Bissiri, Holmes, and Walker}{Bissiri
  et~al.}{2016}]{bissiri2016}
Bissiri, P.~G., C.~Holmes, and S.~Walker (2016).
\newblock {A General Framework for Updating Belief Distributions}.
\newblock {\em Journal of the Royal Statistical Society. Series B
  (Methodological)\/}~{\em 78\/}(5), 1103--1130.

\bibitem[\protect\citeauthoryear{Box}{Box}{1980}]{box1980}
Box, G. E.~P. (1980).
\newblock {Sampling and Bayes' Inference in Scientific Modelling and
  Robustness}.
\newblock {\em Journal of the Royal Statistical Society. Series A
  (General)\/}~{\em 143\/}(4), 383--430.

\bibitem[\protect\citeauthoryear{Dawid, Hartmann, and Sprenger}{Dawid
  et~al.}{2015}]{dawid2015}
Dawid, R., S.~Hartmann, and J.~Sprenger (2015).
\newblock {The No Alternatives Argument}.
\newblock {\em British Journal for the Philosophy of Science\/}~{\em 66\/}(1),
  213--234.

\bibitem[\protect\citeauthoryear{Douven}{Douven}{2016}]{douven2016}
Douven, I. (2016).
\newblock {Explanation, Updating, and Accuracy}.
\newblock {\em Journal of Cognitive Psychology\/}~{\em 28\/}(8), 1004--1012.

\bibitem[\protect\citeauthoryear{Douven and Wenmackers}{Douven and
  Wenmackers}{2017}]{douven2017}
Douven, I. and S.~Wenmackers (2017).
\newblock {Inference to the Best Explanation versus Bayes's Rule in a Social
  Setting}.
\newblock {\em British Journal for the Philosophy of Science\/}~{\em 68\/}(2),
  535--570.

\bibitem[\protect\citeauthoryear{Forster}{Forster}{1995}]{forster1995}
Forster, M.~R. (1995, September).
\newblock Bayes and bust: Simplicity as a problem for a probabilist's approach
  to confirmation.
\newblock {\em British Journal for the Philosophy of Science\/}~{\em 46\/}(3),
  399--424.

\bibitem[\protect\citeauthoryear{Forster and Sober}{Forster and
  Sober}{1994}]{forstersober1994}
Forster, M.~R. and E.~Sober (1994).
\newblock {How To Tell When Simpler, More Unified, or Less Ad Hoc Theories Will
  Provide More Accurate Predictions}.
\newblock {\em The British Journal for the Philosophy of Science\/}~{\em
  45\/}(1), 1--35.

\bibitem[\protect\citeauthoryear{Gelman and Shalizi}{Gelman and
  Shalizi}{2013}]{gelmanshalizi2013}
Gelman, A. and C.~R. Shalizi (2013).
\newblock {Philosophy and the Practice of Bayesian Statistics}.
\newblock {\em British Journal of Mathematical and Statistical
  Psychology\/}~{\em 66}, 8--38.

\bibitem[\protect\citeauthoryear{Gneiting and Raftery}{Gneiting and
  Raftery}{2007}]{gneiting2007}
Gneiting, T. and A.~E. Raftery (2007).
\newblock {Strictly Proper Scoring Rules, Prediction, and Estimation}.
\newblock {\em Journal of the American Statistical Association\/}~{\em
  102\/}(477), 359--378.

\bibitem[\protect\citeauthoryear{Greaves and Wallace}{Greaves and
  Wallace}{2006}]{greaveswallace2006}
Greaves, H. and D.~Wallace (2006).
\newblock {Justifying conditionalization: Conditionalization maximizes
  epistemic utility}.
\newblock {\em Mind\/}~{\em 115\/}(459), 607--632.

\bibitem[\protect\citeauthoryear{Gr{\"u}nwald and van Ommen}{Gr{\"u}nwald and
  van Ommen}{2017}]{grunwald2017}
Gr{\"u}nwald, P. and T.~van Ommen (2017).
\newblock {Inconsistency of Bayesian Inference for Misspecified Linear Models,
  and a Proposal for Repairing It}.
\newblock {\em Bayesian Analysis\/}~{\em 12\/}(4), 1069--1103.

\bibitem[\protect\citeauthoryear{Jeffrey}{Jeffrey}{1983}]{jeffrey1983}
Jeffrey, R. (1983).
\newblock {\em The Logic of Decision\/} (Second ed.).
\newblock Cambridge University Press, Cambridge.

\bibitem[\protect\citeauthoryear{Joyce}{Joyce}{1998}]{joyce1998}
Joyce, J. (1998).
\newblock {A Non-Pragmatic Vindication of Probabilism}.
\newblock {\em Philosophy of Science\/}~{\em 65\/}(4), 575--603.

\bibitem[\protect\citeauthoryear{Joyce}{Joyce}{2009}]{joyce2009}
Joyce, J. (2009).
\newblock {Accuracy and Coherence: Prospects for an Alethic Epistemology of
  Partial Belief}.
\newblock In F.~Huber and C.~Schmidt-Petri (Eds.), {\em Degrees of Belief}.
  Synthese.

\bibitem[\protect\citeauthoryear{Key, Pericchi, and Smith}{Key
  et~al.}{1999}]{key1999}
Key, J.~T., L.~R. Pericchi, and A.~F.~M. Smith (1999).
\newblock {Bayesian Model Choice: What and Why?}
\newblock In J.~M. Bernardo, J.~O. Berger, A.~P. Dawid, and A.~F.~M. Smith
  (Eds.), {\em Bayesian Statistics 6}, pp.\  343--370. Oxford: Oxford
  University Press.

\bibitem[\protect\citeauthoryear{Kopytov and Medvedev}{Kopytov and
  Medvedev}{1996}]{kopytov1996}
Kopytov, V.~M. and N.~Y. Medvedev (1996).
\newblock {\em Right-Ordered Groups}.
\newblock Siberian School of Algebra and Logic. Springer.

\bibitem[\protect\citeauthoryear{Leitgeb and Pettigrew}{Leitgeb and
  Pettigrew}{2010}]{leitgeb2010}
Leitgeb, H. and R.~Pettigrew (2010).
\newblock {An Objective Justification of Bayesianism II: The Consequences of
  Minimizing Inaccuracy}.
\newblock {\em Philosophy of Science\/}~{\em 77}, 236--272.

\bibitem[\protect\citeauthoryear{Levinstein}{Levinstein}{2012}]{levinstein2012}
Levinstein, B.~A. (2012).
\newblock {Leitgeb and Pettigrew on Accuracy and Updating}.
\newblock {\em Philosophy of Science\/}~{\em 79\/}(3), 413--424.

\bibitem[\protect\citeauthoryear{Myrvold}{Myrvold}{2016}]{myrvold2016}
Myrvold, W. (2016).
\newblock {On the Evidential Import of Unification}.
\newblock Unpublished manuscript.

\bibitem[\protect\citeauthoryear{Pettigrew}{Pettigrew}{2016}]{pettigrew2016}
Pettigrew, R. (2016).
\newblock {\em {Accuracy and the Laws of Credence}}.
\newblock Oxford University Press.

\bibitem[\protect\citeauthoryear{Predd, Seiringer, Lieb, Osherson, Poor, and
  Kulkarni}{Predd et~al.}{2009}]{predd2009}
Predd, J.~B., R.~Seiringer, E.~H. Lieb, D.~N. Osherson, H.~V. Poor, and S.~R.
  Kulkarni (2009).
\newblock {Probabilistic Coherence and Proper Scoring Rules}.
\newblock {\em IEEE Transactions on Information Theory\/}~{\em 55\/}(10),
  4786--4792.

\bibitem[\protect\citeauthoryear{Schupbach}{Schupbach}{2018}]{schupbach2018}
Schupbach, J.~N. (2018).
\newblock {Robustness Analysis as Explanatory Reasoning}.
\newblock {\em British Journal for the Philosophy of Science\/}~{\em 69\/}(1),
  275--300.

\bibitem[\protect\citeauthoryear{Shaffer}{Shaffer}{2001}]{shaffer2001}
Shaffer, M.~J. (2001).
\newblock {Bayesian Confirmation of Theories That Incorporate Idealizations}.
\newblock {\em Philosophy of Science\/}~{\em 68\/}(1), 36--52.

\bibitem[\protect\citeauthoryear{Sober}{Sober}{2015}]{sober2015}
Sober, E. (2015).
\newblock {\em Ockham's Razors: A User's Manual}.
\newblock Cambridge University Press.

\bibitem[\protect\citeauthoryear{Sprenger}{Sprenger}{2009}]{sprenger2009}
Sprenger, J. (2009).
\newblock {Statistics Between Inductive Logic and Empirical Science}.
\newblock {\em Journal of Applied Logic\/}~{\em 7\/}(2), 239--250.

\bibitem[\protect\citeauthoryear{Sprenger}{Sprenger}{forthcoming}]{sprenger2017}
Sprenger, J. (forthcoming).
\newblock {Conditional Degree of Belief}.
\newblock To appear in \emph{Philosophy of Science}.

\bibitem[\protect\citeauthoryear{Tanton}{Tanton}{2005}]{tanton2005}
Tanton, J. (2005).
\newblock {\em Encyclopedia of Mathematics}.
\newblock Science Encyclopedia. Facts on File.

\bibitem[\protect\citeauthoryear{Vassend}{Vassend}{2019a}]{vassend2019b}
Vassend, O.~B. (2019a).
\newblock {A Verisimilitude Framework for Inductive Inference, with an
  Application to Phylogenetics}.
\newblock To appear in \emph{British Journal for the Philosophy of Science}.

\bibitem[\protect\citeauthoryear{Vassend}{Vassend}{2019b}]{vassend2019a}
Vassend, O.~B. (2019b).
\newblock {New Semantics for Bayesian Inference: The Interpretive Problem and
  Its Solutions}.
\newblock To appear in \emph{Philosophy of Science}.

\bibitem[\protect\citeauthoryear{Walker}{Walker}{2013}]{walker2013}
Walker, S.~G. (2013).
\newblock {Bayesian Inference with Misspecified Models}.
\newblock {\em Journal of Statistical Planning and Inference\/}~{\em
  143\/}(10), 1621--1633.

\bibitem[\protect\citeauthoryear{Zhang}{Zhang}{2006}]{zhang2006}
Zhang, T. (2006).
\newblock {From e-Entropy to KL-Entropy: Analysis of Minimum Information
  Complexity Density Estimation}.
\newblock {\em The Annals of Statistics\/}~{\em 34\/}(5), 2180--2210.

\end{thebibliography}
\end{document}